\documentclass[aps,prb,superscriptaddress,unsortedaddress,twocolumn]{revtex4}
\usepackage{bm}
\usepackage{amsmath}
\usepackage{amsfonts}%
\usepackage{amssymb}%
\usepackage{graphicx}

\begin{document}

\title{Coupling atomistic and continuum hydrodynamics through a mesoscopic model: application to liquid water}
\author{Rafael Delgado-Buscalioni}
\email[]{rafael.delgado@uam.es}
\affiliation{Departamento F\'{\i}sica Te\'orica de la Materia Condensada,
Universidad Aut\'onoma de Madrid, Campus de Cantoblanco, E-28049 Madrid, Spain}
\author{Kurt Kremer}
\email[]{kremer@mpip-mainz.mpg.de}
\affiliation{Max-Planck-Institut f\"ur Polymerforschung, Ackermannweg 10, D-55128 Mainz, Germany}
\author{Matej Praprotnik}
\email[]{praprot@cmm.ki.si}
\altaffiliation{On leave from the National Institute of Chemistry, Hajdrihova 19,
                 SI-1001 Ljubljana, Slovenia}
\affiliation{Max-Planck-Institut f\"ur Polymerforschung, Ackermannweg 10, D-55128 Mainz, Germany}

\begin{abstract}

 We have conducted a triple-scale simulation of liquid water by
 concurrently coupling atomistic, mesoscopic, and continuum
 models of the liquid. The presented triple-scale hydrodynamic solver 
 for molecular liquids enables the
 insertion of large molecules into the atomistic domain through a
 mesoscopic region. We show that the triple-scale scheme is
robust against the details of the mesoscopic model owing to the conservation
of linear momentum by the adaptive resolution forces. Our multiscale
approach is designed for molecular simulations of open domains with
relatively large molecules, either in the grand canonical ensemble or
under non-equilibrium conditions.
\end{abstract}


\maketitle

\section{Introduction}
Many relevant properties of condensed matter require understanding how
the physics at the nanoscale (nm and ns) builds up or intertwines with
structures and processes on the microscale ($\mu$m and $\mu$s) and
beyond.  The so called {\em multiscale modeling} techniques have been
rapidly evolving during the last decade to bridge this gap.  The
``multiple-scale'' problem is common to many different disciplines,
and a variety of multiscale models are being designed to tackle 
different scenarios either in solids\cite{Broughton:1999,Rottler:2002}
or soft matter\cite{Praprotnik:2008, Koumoutsakos:2005, Malevanets:2000, Donev:2008}. 
The main objective of multiscale modeling of complex fluids
is to study the effect
of large and slow flow scales on the structure and dynamics of complex
molecules (e.g. polymers, proteins), or complex interactions
(e.g. liquid-solid interfaces, wetting fronts, structure formation,
etc.).  In this contexts, multiscale modeling is usually based
on {\em domain decomposition}: a small part of the system [$O(10$nm$)$] 
is solved using fully fledged (classical mechanics)
atomistic detail and it is coupled to a (much larger) outer domain,
described by a coarse-grained (either particle or continuum)
model. The central idea of these ``dual-scale'' methods 
is to solve large and slow processes using a computationally low
demanding description, while retaining an atomistic detail only where necessary. 

Another important application of domain decomposition is the study of
open systems.  Indeed, in general, a dynamic coupling based on domain
decomposition requires to be able to ``open up'' a molecular dynamics
(MD) region, in the sense that mass, momentum and energy should be
exchanged with the exterior in a physically sound way. A
formulation for flux exchange across open
boundaries in particle systems is already available
\cite{Flekkoy:2005}, and was shown to allow for MD simulations in
different type of thermodynamic ensembles.  The triple-scale scheme
presented in this work is equipped with this idea, which permits to study the
dynamics of confined (yet open) molecular systems, which evolve
towards the grand canonical equilibrium ensemble (see e.g.
\cite{Faraudo-prl}). In passing we note that the existing Monte Carlo \cite{Frenkel.book} (MC) 
or hybrid MC-MD algorithms \cite{lynch97} for the grand canonical ensemble  can only 
provide restricted dynamical information of the system.

The first class of schemes based on domain decomposition to appear
were based on particle-continuum coupling (see \cite{Koumoutsakos:2005}
for a review).  In particular, unsteady flow can be solved by hybrids
based on exchanging the momentum flux across the interface (H) between
a molecular dynamics (MD) domain and a continuum fluid dynamics (CFD)
solver. One of these schemes (HybridMD) implements the open-boundary
formulation \cite{Flekkoy:2005}, which allows to open up the MD domain
and include mass and energy exchanges across the MD-CFD interface (H).
This requires consideration of the proper hydrodynamic fluctuations at
the CFD domain \cite{hmd_prl06, hmd_pre07}. However, the particle
insertion used by original ``open MD'' scheme was restricted to small
solvent molecules, such as argon or water \cite{note_usher}, due to
the large steric hindrance of any atomistic complex molecule description.

More recently, another type of domain decomposition, based on
particle-particle coupling, was developed. The Adaptive Resolution
Scheme (AdResS) \cite{Praprotnik:2008,Praprotnik:2005:4} couples a coarse-grained
particle model with its corresponding atomistic description. 
To do so, the number of degrees of freedom of the
molecules is adapted (reduced/increased) as molecules move across a
``transition'' layer where the all-atom explicit model ($ex$) and the coarse-grained
$(cg)$ model are gradually switched on/off, through a hybrid model
($hyb$).  The great benefit of AdResS resides in making feasible the gradual (on-the-fly)
transition of a complex molecule description: from a coarse-grained
potential with soft intermolecular interactions to an atomistic one, with
the whole set of hard-cores.

We realized that taking the advantages of HybridMD and AdResS should then have
a symbiotic effect, potentially solving most of the limitations of
each method. In a recent article \cite{Delgado:2008} we started to
explore in such direction and applied the combined AdResS-HybridMD
model to a liquid of simple tetrahedral molecules.  By performing
molecule insertion within the $cg$ domain, the combined scheme enables
to simulate an open MD system and couple its dynamics to a continuum
flow description of the outer region.  However, the coupling strategy
used in Ref. \cite{Delgado:2008} does not avoid some drawbacks already
present in the original setup of the 
AdResS scheme\cite{Praprotnik:2005:4} (methodological 
improvements to overcome this limitations are underway). 
In particular, precise mapping of
structural and dynamical properties of the $cg$ and $hyb$ molecules
\cite{Reith:2003, Junghans:2008,Tschop:1998,Tschop:1998:2} were still required. Hence, any
simulation exploring a new thermodynamic state require
new calibrations of the coarse-grained models (in practice, each
simulation is restricted to sample one single thermodynamic
state). 

In the present work we show that the coupling geometry can be modified
to yield a more flexible and robust AdResS-HybridMD scheme.  This new
implementation avoids the burden associated to the fine tuning of
coarse-grained layers, thus relieving a great deal of the specificity
of the coarse-grained model. This is important not only from the
computational standpoint, but also because it opens a route to
consider processes following a thermodynamic path. A general completion of this
route requires the inclusion of the energy exchange into the combined
scheme, and some possible solutions are hereby suggested. Section
\ref{s2} briefly introduces the (dual-scale) hybrid models
(HybridMD and AdResS). Coupling strategies are explained in Sec. (\ref{s3}) and
simulations and results are presented in Secs. (\ref{s4}) and
(\ref{s5}). Some conclusions are given in (\ref{s6}).
				   
\section{Hybrid models}
\label{s2}
\subsection{Particle-continuum hybrid (HybridMD)}

The HybridMD \cite{hmd_prl06,hmd_pre07} scheme couples the
hydrodynamic of a particle region, here called ``molecular dynamics''
(MD) domain, with a continuum fluid dynamic description (CFD) of the
external fluid. The essential quantity exchanged between CFD and MD is
the momentum flux across the MD-CFD interface H (see Fig. 1); which can be casted as
${\bf J}_H\cdot {\bf n}$, where ${\bf J}_H$ is the
local pressure tensor and ${\bf n}$ is the unit vector normal to the
H-interface, whose area is $A$.  The momentum flux is transferred to
the MD domain by imposing an external force ${\bf F}^{ext}=A{\bf
  J}_H\cdot {\bf n}$ to a particle buffer 
(the overlapping domain B in Fig. 1) adjacent to
the MD domain.  Molecules are free to cross the H-interface, from or
towards the buffer, but once in B, each molecule $i$ feels an external
``hydrodynamic'' force distributed according to ${\bf F}_i^{ext}=
g(x_i){\bf F}^{ext}/\sum_{i\in B} g(x_i)$, where $x$ is the coordinate
normal to H. Several options can be chosen for the distribution
function \cite{kotsalis:045701}, we used a step function $g(x)=\Theta(x-x_o)$,
as in Ref. \cite{hmd_pre07}.

Through the interface H, the CFD domain receives exactly the same
amount of momentum as the MD system does, but in opposite direction, thus
ensuring conservation. Also, the particle's mass crossing H is injected
into the CFD domain, via a relaxation procedure \cite{hmd_prl06}.
This means that conservation of total mass and momentum only apply to
the system MD$+$CFD.  In other words, strictly speaking B is not part
of the total system \cite{hmd_pre07} and can be thought of as a particle
reservoir where we apply flux boundary conditions to the MD system.
Finally, we note that the only ``microscopic'' information required at the CFD 
level is the equation of state and viscosities of the atomistic fluid (Here, 
we assume that the constitutive relation is known).

\subsection{Adaptive Resolution Scheme (AdResS)}

As stated, the AdResS scheme couples an atomistic domain and a
coarse-grained particle region by adapting, on-the-fly, the molecular
description of those molecules moving across both domains.
This is clearly illustrated in Fig. 1b,c, where the
atomistic domain is labeled as $ex$, the coarse-grained region as
$cg$ and  $hyb$ denotes the transition region between both domains.
In this transition regime, the force acting on a molecule
is an hybrid of the explicit and coarse-grained forces,
\begin{equation}
{\bf F}_{\alpha\beta}=w(X_{\alpha})w(X_{\beta}){\bf
F}^{atom}_{\alpha\beta} + [1-w(X_{\alpha})w(X_{\beta})]{\bf
F}^{cm}_{\alpha\beta},\label{eq:adress}   
\end{equation}
where ${\bf F}_{\alpha\beta}$ is the intermolecular force acting
between centers of mass of molecules $\alpha$ and $\beta$, placed at
$x=X_{\alpha}$ and $X_{\beta}$ in the coupling coordinate, ${\bf
 F}^{atom}_{\alpha\beta}$ is the sum of all pair intermolecular atom
interactions between explicit atoms of the molecules $\alpha$ and
$\beta$, ${\bf F}^{cm}_{\alpha\beta}=-\nabla_{\alpha\beta} U^{cm}$ is
the corresponding coarse-grained force acting on the center of mass 
of the molecule (cm), and $w$ is the weighting function
determining the degree of resolution of the molecules. The value $w=1$ corresponds
to the $ex$ region, $w=0$ to the $cg$ region, while values $0<w<1$
corresponds to hybrid ($hyb$) models. In this study we used the same functional
form of $w$ as in Ref.\cite{Delgado:2008}.

Each time a molecule leaves (or enters) the coarse-grained region it
gradually gains (or loses) its equilibrated vibrational and rotational
DoFs while retaining its linear momentum. Note that the change in
resolution carried out by AdResS is not time-reversible as a given
$cg$ molecule corresponds to many orientations and configurations of
the corresponding $ex$ molecule. Since time reversibility is essential
for energy conservation\cite{Praprotnik:2005}, AdResS does not
conserve energy. In particular, the force in Eq. \ref{eq:adress} is in
general not conservative in the $hyb$ region (i.e., in general $\oint
{\bf F}_{\alpha\beta}\cdot{\bf dr}\ne
0$)\cite{Praprotnik:2007:1,DelleSite:2007}. Hence, to supply or remove
the latent heat associated with the switch of resolution we employ a
standard DPD thermostat\cite{Soddemann:2003,Praprotnik:2007} acting at
the $cg$ and $hyb$ regions.  Note that the thermostat forces 
do not enter into the AdResS interpolating scheme, Eq. (\ref{eq:adress}). 
Instead they are added to the AdResS\cite{Praprotnik:2005:4}.

\section{Coupling strategies}
\label{s3}
The AdResS-HybridMD scheme can be applied in several contexts. 
In general terms, the scheme solves the
hydrodynamic coupling of a MD system with external (CFD) flow. This
goal requires hydrodynamic consistency (e.g. momentum conservation).
Also, a reduced, but yet important, application of the combined scheme
consists on the study of the equilibrium molecular dynamics of open
systems with relatively large molecules. For this sake thermodynamic
consistency is required: in particular sampling the grand-canonical
ensemble requires proper mass fluctuations across the simulation
boundaries. Confined systems are a relevant example of this sort of
application (see e.g. Ref. \cite{Faraudo-prl}), for which the role of CFD
domain can be simplified to just provide the external pressure (and
temperature) of the external mass reservoir.

In the same way, the geometry used for the AdResS-HybridMD coupling
allows for a certain flexibility; the key issue being the location of
the interface H connecting the MD and CFD domains (see Fig. 1).  In a
previous work \cite{Delgado:2008} we proposed to place the H interface
within the coarse-grained domain (see Fig. 1a). This setup is useful
to include hydrodynamics in simulations requiring a 
particle-based multiscale description (e.g. using AdResS).  An example of
application could be the study of the dynamics of ions nearby a
charged surface under flow conditions; here the $ex$ layer would
describe the physics near and at the surface, while away from it, charged $cg$ particles
will eventually transfer the external flow from the outer CFD region.

On the other hand, many studies are focused on the atomistic region and do
not really require a particle-based multiscale description around it.  In
these cases, a more logical setup is to place the interface H within
the atomistic domain $ex$ (see Fig. 1b,c). In this paper we explore the
benefits of this second coupling strategy which, as stated, might open
a route to consistently couple the hydrodynamics and thermodynamics of
qualitatively different levels of description,

\subsection{Coarse-grained buffer}
\label{cgb}
A detailed description of this setup which locates the hybrid
interface H into the $cg$ domain (see Fig. 1a), can be found in
Ref. \cite{Delgado:2008}; we now briefly summarize its requirements.
In this setup the original AdResS scheme is implemented {\em inside} the the MD
domain. Thus, the first requirements of this coupling geometry are
demanded by AdResS \cite{Praprotnik:2005}.  In particular, to
guarantee a similar fluid structure along the AdResS layers, one needs
to calibrate the radial distribution function and equation of state of
the coarse-grained model $cg$ and $hyb$ so as to fit the atomistic
fluid values. Extra pressure corrections require also the same type of
calibrations for the intermediate hybrid model $w=1/2$.  On the other
hand \cite{Delgado:2008}, hydrodynamic consistency demands to take
care of the viscosities of the $cg$ and $hyb$ models, and fit them to
the $ex$ fluid value \cite{Junghans:2008} (as shown in
Ref. \cite{Delgado:2008} this does not
guarantees a perfect fit of the diffusion
coefficients). In summary, for each thermodynamic state considered,
one needs to perform the following calibration steps:

\begin{enumerate}
\item Calibrate the effective potential $U^{cm}_{w=0}$ of the $cg$ model  \cite{Reith:2003} 
so as to fit the center-of-mass radial distribution function (RDF$_{cm}$) and pressure
of the $ex$ model \cite{Praprotnik:2008} (solid line in the inset of Fig. \ref{rdf}).
\item The interface pressure correction \cite{water2} is used to
suppress density oscillations in the $hyb$ layer.  This requires
the calibration of the effective potential $U^{cm}_{w=0.5}$ for the
hybrid model with $w=1/2$ (dashed line in the inset of Fig. \ref{rdf}).
\item Measure the viscosity of the $cg$ model. Then
calibrate the transverse friction coefficient of the transverse DPD
thermostat \cite{Junghans:2008} so as to match $\eta_{cg}$ and
$\eta_{ex}$.  This requires several viscosity calculations \cite{Delgado:2008}.
\end{enumerate}

\subsection{Adaptive resolution buffer}

The setup discussed in this work consists on placing the hybrid
interface H inside the atomistic domain (see Fig. 1b,c). In this setup AdResS
works {\em inside the buffer}. The MD region
is thus purely atomistic and this fact brings about an important benefit:
around the hybrid interface H, all relevant fluid properties are
properly defined. These include the fluid transport coefficients,
energy and pressure equation of states and the corresponding mass and
pressure fluctuations. In this context, MD-CFD coupling by HybridMD is
well defined \cite{hmd_prl06}.  On the other side, by placing AdResS
into the particle buffer it behaves as an {\em adaptive resolution buffer}, 
where the atomistic DoF's are gradually inserted into the
MD (atomistic) region of interest.  Note that the $cg$ layer, with soft
interaction potentials, is placed near the buffer end (see Fig. 1b,c), so
the insertion of complex molecules into the system is still an easy
task, using standard insertion routines \cite{Delgado:2003:1}.

The second relevant benefit of this coupling strategy is that it
permits to properly impose the external pressure and stress into MD,
without having to perform a fine tuning of the structural and dynamic
properties of the $cg$ and $hyb$ models. In other words, the
AdResS-HybridMD scheme does not rely on the specificity of the
coarse-grained description anymore.  The reason for this is simple:
the AdResS scheme conserves linear momentum. This means that any
external momentum flux will be properly transferred across the AdResS
layers to the atomistic core. As the external ``hydrodynamic'' force
used in the HybridMD scheme decouples from any intermolecular
interaction (the total force on a molecule at the buffer B is, ${\bf
  F}_{\alpha}={\bf F}_{\alpha}^{ext}+\sum_{\beta} {\bf F}_{\alpha
  \beta}$), the transfer of mechanical variables (pressure and stress)
should be robust against the details of the $cg$ and $hyb$ models.  We
now present simulations to prove this claim.  We will also consider
any effect on the liquid structure near the interface H and check for
proper mass fluctuations across H.

\section{Simulations}
\label{s4}

The present hybrid simulations were implemented for the flexible TIP3P
water model, partly because of the relevance of water and also because
it has well known structural and dynamical
properties\cite{Jorgensen:1983,Neria:1996,Praprotnik:2005:2,fh07}.  In
the remainder of the paper we use reduced Lennard-Jones units
corresponding to the Oxygen atom: mass $m=m_O = 16$ a.u.,
oxygen-oxygen interaction energy $\epsilon=\epsilon_{OO}=0.152073
\mathrm{kcal/mol}$, and diameter $\sigma=\sigma_O=3.1507
\AA$. Simulations were done at ambient temperature, $T=300$K, which
corresponds to $k_B T/\epsilon=3.92$ in reduced units. TIP3P-water
density is $1.02$ gr/cm$^3$, which corresponds to
$\rho=1.20\,m/\sigma^3$. Most particle simulations were done in boxes
of total (MD$+$B) volume $24.5\times 6.18 \times 11.12 \sigma^3$,
although in order check for finite size effects, boxes with larger
dimensions were also used. The volume of the MD domain was
$V=10.5\times 6.18 \times 11.12 \sigma^3$.  and it contained about 865
water molecules.  Long-range electrostatic forces are computed using
the reaction field method, in which all molecules outside a spherical
cavity of a molecular based cutoff radius $R_c=2.86\sigma$ are treated
as a dielectric continuum with a dielectric constant
$\varepsilon_{RF}=80$~\cite{Neumann:1983, Praprotnik:2007:2,water2}.
The finite volume solver for the CFD domain was feeded with the
equation of state for flexible TIP3P water
reported in Ref. \cite{fh07}.  Finally, the microscopic part of the
stress tensor at the hybrid interface H was measured according to the 
mesoscopic approach explained in Ref. \cite{hmd_pre07}; i.e.  by
evaluating velocity gradients from the cell-averaged particle velocities and
using the Newtonian constitutive relation.

\section{Results}
\label{s5}
{\em Coarse-grained buffers}. As expected, results proved the
correct behavior of the combined scheme, both in terms of the
structure and hydrodynamics at the MD domain. However calibration
steps, using the recipe explained in Sec. \ref{cgb}, involved
significant work. Strong corrections of the $cg$ and $hyb$ viscosities
\cite{Junghans:2008} were required (without transverse DPD thermostat,
$\eta_{cg}$ is about five times smaller than $\eta_{ex}$). In these
calibration steps we used HybridMD as a rheometer \cite{hmd_pre07}.

{\em Adaptive resolution buffers}.  In what follows we focus on the
result obtained for the second, much lighter setup. It is important to
stress that these simulations were done using an effective potential
for the $cg$ model which was deliberately not accurately fitted to
reproduce the all-atom RDF$_{cm}$ (see Fig. \ref{rdf}).  Moreover,
steps $2$ and $3$ of the protocol of Sec. \ref{cgb} were also avoided,
meaning that the shear viscosities at the $cg$ and $hyb$ layers result
in much smaller values than the atomistic ($ex$) one.

\subsection{Liquid structure}
We first compare, in Fig. \ref{rdf}, the local structure
of the liquid inside the MD region of the triple-scale model with that
obtained from all-atom simulations within periodic boundaries. The
agreement is perfect, indicating that the unfitted liquid structure in
the buffer (dashed line in Fig. \ref{rdf}) does not affect the
proper liquid structure inside the interest (MD) region.

\subsection{Hydrodynamics}
The hydrodynamic behavior of the triple-scale scheme was tested
considering both steady and unsteady flows.  Figure \ref{shear}
shows the density and velocity profiles in the particle
region (MD$+$B) obtained at the steady state of a simple Couette
flow. The density (and temperature, not shown) profile inside the MD region 
is flat and confirm that the triple-scale scheme furnishes an homogeneous equilibrated liquid bulk. 
The expected linear velocity profile inside the MD domain indicates that the
transverse momentum is correctly transferred across the triple-scale
fluid model. We also conducted simulations of Stokes flow 
(an oscillatory shear flow
driven by the oscillatory motion of a wall along its plane
direction). Figure \ref{osc_shear} shows the velocity in one of the
MD-cells of the system corresponding to a flow with a period 300$\tau$.  The
hybrid result perfectly agrees with with the deterministic solution of
the Navier-Stokes equation (red solid line in Fig. \ref{osc_shear}).
The large viscosity of liquid water induces a fast transfer of momenta
across the buffers. Hence, even for faster shear rates, no trace of
phase delay between the MD and CFD velocities is observed
\cite{Delgado:2008}, (see the inset of Fig. \ref{osc_shear}).

\subsection{Mass fluctuations}
One of the interesting properties of the AdResS-HybridMD approach is that the
MD region becomes an open system, which exchanges mass with its
surroundings. As stated before, to that end 
it is quite important to check that mass fluctuation
across the MD border (H) is consistent with the grand canonical prescription.
We measured the mass variance inside the MD domain and
compared it with the grand canonical (GC) result, $\mathrm{Var}[\rho] =
\rho k_B T/ (V c_T^2)$, where $V$ is the system's volume and $c_T^2
=(\partial P/\partial \rho)_T$ is the squared isothermal sound
velocity (related to the isothermal compressibility, $\beta_T= (c_T^2
\rho)^{-1}$).  Taking the sound velocity for the flexible TIP3P water
reported in Ref. \cite{fh07}, $c_T = 7.38 (\epsilon/m)^{1/2}$ and the
mass density $\rho=1.20 m/\sigma^3$ (recall that $m=m_O$), the GC
prediction for $V=3.5\times 6.18\times 11.12
\sigma^3$ is $\mathrm{Var}[\rho]=0.0187$.  Inside the MD domain we
obtained $\mathrm{Var}[\rho]=0.020\pm 0.002$ within different slices
of the same volume. This is a quite good agreement, considering the
smallness of $V$.  In a larger volume $V=10.5\times 6.18\times 11.12
\sigma^3$, the triple-scale result $\mathrm{Var}[\rho]=0.011\pm 0.005$ is
even closer to the GC prediction $\mathrm{Var}[\rho]=0.0108$.

\subsection{Energy}
Although in this work we do not solve the energy exchange across H, we
are in position to provide some arguments indicating that this task is
solvable using the {\em adaptive resolution buffer} setup. Energy
exchange requires three properties: $1$) energy should be properly
defined across MD, $2$) the scheme should allow to change the
thermodynamic state of the system and $3$) one should be able to
control the amount of energy per unit time inserted into
MD. Concerning $1$), it is important to stress that by placing the
AdResS scheme inside the buffer one ensures that the energy is a well
defined quantity over the total (MD$+$CFD) system. Also $2$) is
satisfied because the {\em adaptive resolution buffer} setup does not
rely on the specificity coarse-grained model: this means that it should
be possible to change the thermodynamic state of the MD domain (e.g. the
mean temperature) without re-calibrating the $cg$ layer 
(we have tested this in simulations using tetrahedral molecules). Finally,
we believe that $3$) is solvable, but it will require
a further development of the algorithm. 
One possible way could be to implement the flux-boundary
conditions developed in Ref. \cite{Flekkoy:2005}.
We expect to present an algorithm of this sort in
future works.

\section{Conclusions}
\label{s6}
To conclude, we have presented a flexible and robust hybrid scheme for
hydrodynamics of molecular liquid, which combines atomistic, mesoscopic
and continuum models. This triple-scale scheme uses a flux based
particle-continuum hybrid to couple the atomistic core and continuum
sides of the system, while it generalizes the role of the particle
buffer to allow for a gradual change in molecular resolution: from
an all-atom at the core to a mesoscopic one near the buffer end.
Structure and hydrodynamic of the core (MD) region were shown to be
robust against changes in the choice of the mesoscopic model, greatly
reducing calibration burdens. Further extensions to allow for 
energy exchange and multiple species will be explored in future works.

{\bf Acknowledgments}: We thank Anne Dejoan, Christoph Junghans, Burkhard D{\" u}nweg, and 
 Luigi Delle Site for useful discussions. This work is supported in part by the
 Volkswagen foundation. R.D-B acknowledges additional funding from
 Ram\'on y Cajal contract and project FIS2007-65869-C03-01 funded by the
 Spanish government (MEC), and also funding from the "Comunidad de
 Madrid" through the MOSSNOHO project, S-0505/ESP/0299. M.~P. acknowledges
 additional financial support through the project J1-2281 from the Slovenian 
 Research Agency.


\begin{thebibliography}{32}
\expandafter\ifx\csname natexlab\endcsname\relax\def\natexlab#1{#1}\fi
\expandafter\ifx\csname bibnamefont\endcsname\relax
  \def\bibnamefont#1{#1}\fi
\expandafter\ifx\csname bibfnamefont\endcsname\relax
  \def\bibfnamefont#1{#1}\fi
\expandafter\ifx\csname citenamefont\endcsname\relax
  \def\citenamefont#1{#1}\fi
\expandafter\ifx\csname url\endcsname\relax
  \def\url#1{\texttt{#1}}\fi
\expandafter\ifx\csname urlprefix\endcsname\relax\def\urlprefix{URL }\fi
\providecommand{\bibinfo}[2]{#2}
\providecommand{\eprint}[2][]{\url{#2}}

\bibitem[{\citenamefont{Broughton et~al.}(1999)\citenamefont{Broughton,
  Abraham, Bernstein, and Kaxiras}}]{Broughton:1999}
\bibinfo{author}{\bibfnamefont{J.~Q.} \bibnamefont{Broughton}},
  \bibinfo{author}{\bibfnamefont{F.~F.} \bibnamefont{Abraham}},
  \bibinfo{author}{\bibfnamefont{N.}~\bibnamefont{Bernstein}},
  \bibnamefont{and} \bibinfo{author}{\bibfnamefont{E.}~\bibnamefont{Kaxiras}},
  \bibinfo{journal}{Phys. Rev. B} \textbf{\bibinfo{volume}{60}},
  \bibinfo{pages}{2391} (\bibinfo{year}{1999}).

\bibitem[{\citenamefont{Rottler et~al.}(2002)\citenamefont{Rottler, Barsky, and
  Robbins}}]{Rottler:2002}
\bibinfo{author}{\bibfnamefont{J.}~\bibnamefont{Rottler}},
  \bibinfo{author}{\bibfnamefont{S.}~\bibnamefont{Barsky}}, \bibnamefont{and}
  \bibinfo{author}{\bibfnamefont{M.~O.} \bibnamefont{Robbins}},
  \bibinfo{journal}{Phys. Rev. Lett.} \textbf{\bibinfo{volume}{89}},
  \bibinfo{pages}{148304} (\bibinfo{year}{2002}).

\bibitem[{\citenamefont{Praprotnik et~al.}(2008)\citenamefont{Praprotnik,
  Delle~Site, and Kremer}}]{Praprotnik:2008}
\bibinfo{author}{\bibfnamefont{M.}~\bibnamefont{Praprotnik}},
  \bibinfo{author}{\bibfnamefont{L.}~\bibnamefont{Delle~Site}},
  \bibnamefont{and} \bibinfo{author}{\bibfnamefont{K.}~\bibnamefont{Kremer}},
  \bibinfo{journal}{Annu. Rev. Phys. Chem.} \textbf{\bibinfo{volume}{59}},
  \bibinfo{pages}{545} (\bibinfo{year}{2008}).

\bibitem[{\citenamefont{Koumoutsakos}(2005)}]{Koumoutsakos:2005}
\bibinfo{author}{\bibfnamefont{P.}~\bibnamefont{Koumoutsakos}},
  \bibinfo{journal}{Annu.\ Rev.\ Fluid Mech.} \textbf{\bibinfo{volume}{37}},
  \bibinfo{pages}{457} (\bibinfo{year}{2005}).

\bibitem[{\citenamefont{Malevanets and Kapral}(2000)}]{Malevanets:2000}
\bibinfo{author}{\bibfnamefont{A.}~\bibnamefont{Malevanets}} \bibnamefont{and}
  \bibinfo{author}{\bibfnamefont{R.}~\bibnamefont{Kapral}},
  \bibinfo{journal}{J. Chem. Phys.} \textbf{\bibinfo{volume}{112}},
  \bibinfo{pages}{7260} (\bibinfo{year}{2000}).

\bibitem[{\citenamefont{Donev et~al.}(2008)\citenamefont{Donev, Alder, and
  Garcia}}]{Donev:2008}
\bibinfo{author}{\bibfnamefont{A.}~\bibnamefont{Donev}},
  \bibinfo{author}{\bibfnamefont{B.~J.} \bibnamefont{Alder}}, \bibnamefont{and}
  \bibinfo{author}{\bibfnamefont{A.~L.} \bibnamefont{Garcia}},
  \bibinfo{journal}{Phys.\ Rev.\ Lett.} \textbf{\bibinfo{volume}{101}},
  \bibinfo{pages}{075902} (\bibinfo{year}{2008}).

\bibitem[{\citenamefont{Flekkoy et~al.}(2005)\citenamefont{Flekkoy,
  Delgado-Buscalioni, and Coveney}}]{Flekkoy:2005}
\bibinfo{author}{\bibfnamefont{E.~G.} \bibnamefont{Flekkoy}},
  \bibinfo{author}{\bibfnamefont{R.}~\bibnamefont{Delgado-Buscalioni}},
  \bibnamefont{and} \bibinfo{author}{\bibfnamefont{P.~V.}
  \bibnamefont{Coveney}}, \bibinfo{journal}{Phys. Rev. E}
  \textbf{\bibinfo{volume}{72}}, \bibinfo{pages}{026703}
  (\bibinfo{year}{2005}).

\bibitem[{\citenamefont{Faraudo and Bresme}(2004)}]{Faraudo-prl}
\bibinfo{author}{\bibfnamefont{J.}~\bibnamefont{Faraudo}} \bibnamefont{and}
  \bibinfo{author}{\bibfnamefont{F.}~\bibnamefont{Bresme}},
  \bibinfo{journal}{Phys. Rev. Lett.} \textbf{\bibinfo{volume}{92}},
  \bibinfo{pages}{236102} (\bibinfo{year}{2004}).

\bibitem[{\citenamefont{Frenkel and Smith}(2002)}]{Frenkel.book}
\bibinfo{author}{\bibfnamefont{D.}~\bibnamefont{Frenkel}} \bibnamefont{and}
  \bibinfo{author}{\bibfnamefont{B.}~\bibnamefont{Smith}},
  \emph{\bibinfo{title}{Understanding Molecular Simulation: From Algorithms to
  Applications}} (\bibinfo{publisher}{Academic Press, San Diego, 2nd edition},
  \bibinfo{year}{2002}).

\bibitem[{\citenamefont{Lynch and Pettitt}(1997)}]{lynch97}
\bibinfo{author}{\bibfnamefont{G.~C.} \bibnamefont{Lynch}} \bibnamefont{and}
  \bibinfo{author}{\bibfnamefont{B.~M.} \bibnamefont{Pettitt}},
  \bibinfo{journal}{J. Chem. Phys.} \textbf{\bibinfo{volume}{107}},
  \bibinfo{pages}{8594} (\bibinfo{year}{1997}).

\bibitem[{\citenamefont{{De Fabritiis} et~al.}(2006)\citenamefont{{De
  Fabritiis}, Delgado-Buscalioni, and Coveney}}]{hmd_prl06}
\bibinfo{author}{\bibfnamefont{G.}~\bibnamefont{{De Fabritiis}}},
  \bibinfo{author}{\bibfnamefont{R.}~\bibnamefont{Delgado-Buscalioni}},
  \bibnamefont{and} \bibinfo{author}{\bibfnamefont{P.}~\bibnamefont{Coveney}},
  \bibinfo{journal}{Phys. Rev. Lett} \textbf{\bibinfo{volume}{97}},
  \bibinfo{pages}{134501} (\bibinfo{year}{2006}).

\bibitem[{\citenamefont{Delgado-Buscalioni and {De
  Fabritiis}}(2007)}]{hmd_pre07}
\bibinfo{author}{\bibfnamefont{R.}~\bibnamefont{Delgado-Buscalioni}}
  \bibnamefont{and} \bibinfo{author}{\bibfnamefont{G.}~\bibnamefont{{De
  Fabritiis}}}, \bibinfo{journal}{Phys.\ Rev.\ E}
  \textbf{\bibinfo{volume}{76}}, \bibinfo{pages}{036709}
  (\bibinfo{year}{2007}).

\bibitem[{not()}]{note_usher}
\bibinfo{note}{In the original HybridMD scheme molecule insertion is done using
  the {\sc usher} algorithm, which was originally designed for Lennard-Jones
  particles \cite{Delgado:2003:1} and water \cite{water_usher}}.


\bibitem[{\citenamefont{De Fabritiis et~al.}(2007)}]{water_usher}
\bibinfo{author}{\bibfnamefont{G.}~\bibnamefont{{De
  Fabritiis}}}
\bibinfo{author}{\bibfnamefont{R.}~\bibnamefont{Delgado-Buscalioni}}
  \bibnamefont{and} \bibinfo{author}{\bibfnamefont{P.~V.}~\bibnamefont{{Coveney}}} , \bibinfo{journal}{J. Chem. Phys.}
  \textbf{\bibinfo{volume}{121}}, \bibinfo{pages}{12139}
  (\bibinfo{year}{2004}).


\bibitem[{\citenamefont{Praprotnik et~al.}(2005)\citenamefont{Praprotnik,
  Delle~Site, and Kremer}}]{Praprotnik:2005:4}
\bibinfo{author}{\bibfnamefont{M.}~\bibnamefont{Praprotnik}},
  \bibinfo{author}{\bibfnamefont{L.}~\bibnamefont{Delle~Site}},
  \bibnamefont{and} \bibinfo{author}{\bibfnamefont{K.}~\bibnamefont{Kremer}},
  \bibinfo{journal}{J. Chem. Phys.} \textbf{\bibinfo{volume}{123}},
  \bibinfo{pages}{224106} (\bibinfo{year}{2005}).

\bibitem[{\citenamefont{Delgado-Buscalioni
  et~al.}(2008)\citenamefont{Delgado-Buscalioni, Kremer, and
  Praprotnik}}]{Delgado:2008}
\bibinfo{author}{\bibfnamefont{R.}~\bibnamefont{Delgado-Buscalioni}},
  \bibinfo{author}{\bibfnamefont{K.}~\bibnamefont{Kremer}}, \bibnamefont{and}
  \bibinfo{author}{\bibfnamefont{M.}~\bibnamefont{Praprotnik}},
  \bibinfo{journal}{J. Chem. Phys.} \textbf{\bibinfo{volume}{128}},
  \bibinfo{pages}{114110} (\bibinfo{year}{2008}).

\bibitem[{\citenamefont{Reith et~al.}(2003)\citenamefont{Reith, P{\" u}tz, and
  M{\" u}ller-Plathe}}]{Reith:2003}
\bibinfo{author}{\bibfnamefont{D.}~\bibnamefont{Reith}},
  \bibinfo{author}{\bibfnamefont{M.}~\bibnamefont{P{\" u}tz}},
  \bibnamefont{and} \bibinfo{author}{\bibfnamefont{F.}~\bibnamefont{M{\"
  u}ller-Plathe}}, \bibinfo{journal}{J. Comput. Chem.}
  \textbf{\bibinfo{volume}{24}}, \bibinfo{pages}{1624} (\bibinfo{year}{2003}).

\bibitem[{\citenamefont{Junghans et~al.}(2008)\citenamefont{Junghans,
  Praprotnik, and Kremer}}]{Junghans:2008}
\bibinfo{author}{\bibfnamefont{C.}~\bibnamefont{Junghans}},
  \bibinfo{author}{\bibfnamefont{M.}~\bibnamefont{Praprotnik}},
  \bibnamefont{and} \bibinfo{author}{\bibfnamefont{K.}~\bibnamefont{Kremer}},
  \bibinfo{journal}{Soft Matter} \textbf{\bibinfo{volume}{4}},
  \bibinfo{pages}{156} (\bibinfo{year}{2008}).

\bibitem[{\citenamefont{Tsch{\" o}p
  et~al.}(1998{\natexlab{a}})\citenamefont{Tsch{\" o}p, Kremer, Batoulis, B{\"
  u}rger, and Hahn}}]{Tschop:1998}
\bibinfo{author}{\bibfnamefont{W.}~\bibnamefont{Tsch{\" o}p}},
  \bibinfo{author}{\bibfnamefont{K.}~\bibnamefont{Kremer}},
  \bibinfo{author}{\bibfnamefont{J.}~\bibnamefont{Batoulis}},
  \bibinfo{author}{\bibfnamefont{T.}~\bibnamefont{B{\" u}rger}},
  \bibnamefont{and} \bibinfo{author}{\bibfnamefont{O.}~\bibnamefont{Hahn}},
  \bibinfo{journal}{Acta Polym.} \textbf{\bibinfo{volume}{49}},
  \bibinfo{pages}{61} (\bibinfo{year}{1998}{\natexlab{a}}).

\bibitem[{\citenamefont{Tsch{\" o}p
  et~al.}(1998{\natexlab{b}})\citenamefont{Tsch{\" o}p, Kremer, Hahn, Batoulis,
  and B{\" u}rger}}]{Tschop:1998:2}
\bibinfo{author}{\bibfnamefont{W.}~\bibnamefont{Tsch{\" o}p}},
  \bibinfo{author}{\bibfnamefont{K.}~\bibnamefont{Kremer}},
  \bibinfo{author}{\bibfnamefont{O.}~\bibnamefont{Hahn}},
  \bibinfo{author}{\bibfnamefont{J.}~\bibnamefont{Batoulis}}, \bibnamefont{and}
  \bibinfo{author}{\bibfnamefont{T.}~\bibnamefont{B{\" u}rger}},
  \bibinfo{journal}{Acta Polym.} \textbf{\bibinfo{volume}{49}},
  \bibinfo{pages}{75} (\bibinfo{year}{1998}{\natexlab{b}}).

\bibitem[{\citenamefont{Kotsalis et~al.}(2009)\citenamefont{Kotsalis, Walther,
  Kaxiras, and Koumoutsakos}}]{kotsalis:045701}
\bibinfo{author}{\bibfnamefont{E.~M.} \bibnamefont{Kotsalis}},
  \bibinfo{author}{\bibfnamefont{J.~H.} \bibnamefont{Walther}},
  \bibinfo{author}{\bibfnamefont{E.}~\bibnamefont{Kaxiras}}, \bibnamefont{and}
  \bibinfo{author}{\bibfnamefont{P.}~\bibnamefont{Koumoutsakos}},
  \bibinfo{journal}{Phys.\ Rev.\ E} \textbf{\bibinfo{volume}{79}},
  \bibinfo{pages}{045701} (\bibinfo{year}{2009}).

\bibitem[{\citenamefont{Jane\v{z}i\v{c}
  et~al.}(2005)\citenamefont{Jane\v{z}i\v{c}, Praprotnik, and
  Merzel}}]{Praprotnik:2005}
\bibinfo{author}{\bibfnamefont{D.}~\bibnamefont{Jane\v{z}i\v{c}}},
  \bibinfo{author}{\bibfnamefont{M.}~\bibnamefont{Praprotnik}},
  \bibnamefont{and} \bibinfo{author}{\bibfnamefont{F.}~\bibnamefont{Merzel}},
  \bibinfo{journal}{J. Chem. Phys.} \textbf{\bibinfo{volume}{122}},
  \bibinfo{pages}{174101} (\bibinfo{year}{2005}).

\bibitem[{\citenamefont{Praprotnik
  et~al.}(2007{\natexlab{a}})\citenamefont{Praprotnik, Kremer, and
  Delle~Site}}]{Praprotnik:2007:1}
\bibinfo{author}{\bibfnamefont{M.}~\bibnamefont{Praprotnik}},
  \bibinfo{author}{\bibfnamefont{K.}~\bibnamefont{Kremer}}, \bibnamefont{and}
  \bibinfo{author}{\bibfnamefont{L.}~\bibnamefont{Delle~Site}},
  \bibinfo{journal}{J. Phys. A: Math. Theor.} \textbf{\bibinfo{volume}{40}},
  \bibinfo{pages}{F281} (\bibinfo{year}{2007}{\natexlab{a}}).

\bibitem[{\citenamefont{Delle~Site}(2007)}]{DelleSite:2007}
\bibinfo{author}{\bibfnamefont{L.}~\bibnamefont{Delle~Site}},
  \bibinfo{journal}{Phys. Rev. E} \textbf{\bibinfo{volume}{76}},
  \bibinfo{pages}{047701} (\bibinfo{year}{2007}).

\bibitem[{\citenamefont{Soddemann et~al.}(2003)\citenamefont{Soddemann, D{\"
  u}nweg, and Kremer}}]{Soddemann:2003}
\bibinfo{author}{\bibfnamefont{T.}~\bibnamefont{Soddemann}},
  \bibinfo{author}{\bibfnamefont{B.}~\bibnamefont{D{\" u}nweg}},
  \bibnamefont{and} \bibinfo{author}{\bibfnamefont{K.}~\bibnamefont{Kremer}},
  \bibinfo{journal}{Phys. Rev. E} \textbf{\bibinfo{volume}{68}},
  \bibinfo{pages}{046702} (\bibinfo{year}{2003}).

\bibitem[{\citenamefont{Praprotnik
  et~al.}(2007{\natexlab{b}})\citenamefont{Praprotnik, Kremer, and
  Delle~Site}}]{Praprotnik:2007}
\bibinfo{author}{\bibfnamefont{M.}~\bibnamefont{Praprotnik}},
  \bibinfo{author}{\bibfnamefont{K.}~\bibnamefont{Kremer}}, \bibnamefont{and}
  \bibinfo{author}{\bibfnamefont{L.}~\bibnamefont{Delle~Site}},
  \bibinfo{journal}{Phys. Rev. E} \textbf{\bibinfo{volume}{75}},
  \bibinfo{pages}{017701} (\bibinfo{year}{2007}{\natexlab{b}}).

\bibitem[{\citenamefont{Matysiak et~al.}(2008)\citenamefont{Matysiak, Clementi,
  Praprotnik, Kremer, and {Delle Site}}}]{water2}
\bibinfo{author}{\bibfnamefont{S.}~\bibnamefont{Matysiak}},
  \bibinfo{author}{\bibfnamefont{C.}~\bibnamefont{Clementi}},
  \bibinfo{author}{\bibfnamefont{M.}~\bibnamefont{Praprotnik}},
  \bibinfo{author}{\bibfnamefont{K.}~\bibnamefont{Kremer}}, \bibnamefont{and}
  \bibinfo{author}{\bibfnamefont{L.}~\bibnamefont{{Delle Site}}},
  \bibinfo{journal}{J. Chem. Phys.} \textbf{\bibinfo{volume}{128}},
  \bibinfo{pages}{024503} (\bibinfo{year}{2008}).

\bibitem[{\citenamefont{Delgado-Buscalioni and Coveney}(2003)}]{Delgado:2003:1}
\bibinfo{author}{\bibfnamefont{R.}~\bibnamefont{Delgado-Buscalioni}}
  \bibnamefont{and} \bibinfo{author}{\bibfnamefont{P.~V.}
  \bibnamefont{Coveney}}, \bibinfo{journal}{J. Chem. Phys.}
  \textbf{\bibinfo{volume}{119}}, \bibinfo{pages}{978} (\bibinfo{year}{2003}).

\bibitem[{\citenamefont{Jorgensen et~al.}(1983)\citenamefont{Jorgensen,
  Chandrasekhar, Madura, Impey, and Klein}}]{Jorgensen:1983}
\bibinfo{author}{\bibfnamefont{W.~L.} \bibnamefont{Jorgensen}},
  \bibinfo{author}{\bibfnamefont{J.}~\bibnamefont{Chandrasekhar}},
  \bibinfo{author}{\bibfnamefont{J.~D.} \bibnamefont{Madura}},
  \bibinfo{author}{\bibfnamefont{R.~W.} \bibnamefont{Impey}}, \bibnamefont{and}
  \bibinfo{author}{\bibfnamefont{M.~L.} \bibnamefont{Klein}},
  \bibinfo{journal}{J. Chem. Phys.} \textbf{\bibinfo{volume}{79}},
  \bibinfo{pages}{926} (\bibinfo{year}{1983}).

\bibitem[{\citenamefont{Praprotnik and
  Jane\v{z}i\v{c}}(2005)}]{Praprotnik:2005:2}
\bibinfo{author}{\bibfnamefont{M.}~\bibnamefont{Praprotnik}} \bibnamefont{and}
  \bibinfo{author}{\bibfnamefont{D.}~\bibnamefont{Jane\v{z}i\v{c}}},
  \bibinfo{journal}{J. Chem. Phys.} \textbf{\bibinfo{volume}{122}},
  \bibinfo{pages}{174103} (\bibinfo{year}{2005}).

\bibitem[{\citenamefont{{De Fabritiis} et~al.}(2007)\citenamefont{{De
  Fabritiis}, Serrano, Delgado-Buscalioni, and Coveney}}]{fh07}
\bibinfo{author}{\bibfnamefont{G.}~\bibnamefont{{De Fabritiis}}},
  \bibinfo{author}{\bibfnamefont{M.}~\bibnamefont{Serrano}},
  \bibinfo{author}{\bibfnamefont{R.}~\bibnamefont{Delgado-Buscalioni}},
  \bibnamefont{and} \bibinfo{author}{\bibfnamefont{P.~V.}
  \bibnamefont{Coveney}}, \bibinfo{journal}{Phys.\ Rev.\ E}
  \textbf{\bibinfo{volume}{75}}, \bibinfo{pages}{026307}
  (\bibinfo{year}{2007}).

\bibitem[{\citenamefont{Neria et~al.}(1996)\citenamefont{Neria, Fischer, and
  Karplus}}]{Neria:1996}
\bibinfo{author}{\bibfnamefont{E.}~\bibnamefont{Neria}},
  \bibinfo{author}{\bibfnamefont{S.}~\bibnamefont{Fischer}}, \bibnamefont{and}
  \bibinfo{author}{\bibfnamefont{M.}~\bibnamefont{Karplus}},
  \bibinfo{journal}{J.\ Chem.\ Phys.} \textbf{\bibinfo{volume}{105}},
  \bibinfo{pages}{1902} (\bibinfo{year}{1996}).

\bibitem[{\citenamefont{Neumann}(1983)}]{Neumann:1983}
\bibinfo{author}{\bibfnamefont{M.}~\bibnamefont{Neumann}},
  \bibinfo{journal}{Mol. Phys.} \textbf{\bibinfo{volume}{50}},
  \bibinfo{pages}{841} (\bibinfo{year}{1983}).

\bibitem[{\citenamefont{Praprotnik
  et~al.}(2007{\natexlab{c}})\citenamefont{Praprotnik, Matysiak, {Delle Site},
  Kremer, and Clementi}}]{Praprotnik:2007:2}
\bibinfo{author}{\bibfnamefont{M.}~\bibnamefont{Praprotnik}},
  \bibinfo{author}{\bibfnamefont{S.}~\bibnamefont{Matysiak}},
  \bibinfo{author}{\bibfnamefont{L.}~\bibnamefont{{Delle Site}}},
  \bibinfo{author}{\bibfnamefont{K.}~\bibnamefont{Kremer}}, \bibnamefont{and}
  \bibinfo{author}{\bibfnamefont{C.}~\bibnamefont{Clementi}},
  \bibinfo{journal}{J. Phys.: Condens. Matter} \textbf{\bibinfo{volume}{19}},
  \bibinfo{pages}{292201} (\bibinfo{year}{2007}{\natexlab{c}}).

\end{thebibliography}

\newpage

\begin{figure}[!ht]
\includegraphics[width=\linewidth]{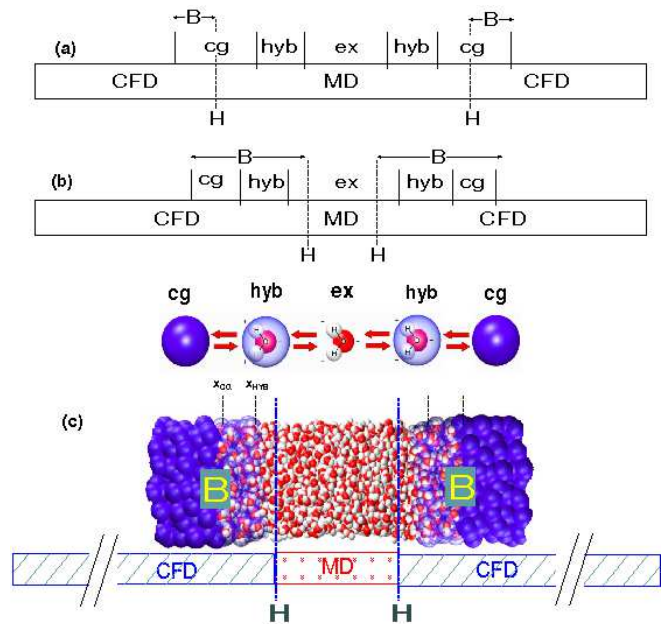}
\caption{Coupling strategies for the AdResS-HybridMD scheme: (a) {\em
    Coarse-grained buffer} and (b), (c) {\em Adaptive resolution
    buffer}.  In each figure, the bottom panel depicts the
  decomposition of the whole system (MD$+$CFD); where MD stands for
  the molecular dynamics region, surrounded by continuum fluid
  dynamics (CFD) domains solved via the finite volume method
  \cite{fh07}.  The MD-CFD coupling is solved by the HybridMD
  scheme\cite{hmd_prl06}, based on the exchange of momentum flux
  across the interface H.  Pressure and stress are imposed into MD
  via external forces acting on particles at the buffers B.  The
  AdResS scheme \cite{Praprotnik:2008} (see the middle figure)
  gradually adapts the atomic resolution of the molecules: from
  all-atom ($ex$) to coarse-grained ($cg$) descriptions, passing
  through a hybrid ($hyb$) model.  AdResS and HybridMD can be combined
  in two ways depending on location of the interface H: either using a
  {\em Coarse-grained buffer} (a), see Ref. \cite{Delgado:2008}; or 
an {\em Adaptive resolution buffer} (b) and (c), explored in this work.  
Figure (c) is an illustration of this triple-scale scheme for liquid water.  The
  hydrodynamic coupling is made along $x$ direction, (finite volume
  cells are $\Delta x = 3.5\sigma$ wide) and the system is periodic in
  the orthogonal directions.  The atomic resolution of water molecules
  is gradually switched on as they move across the buffer, which is $7
  \sigma$ long. The $hyb$ region is $3.5 \sigma$ and its distance to H
  is about 1$\sigma$.  A standard DPD thermostat at $T=300K$ is used
  for the $cg$ and $hyb$ domains, with a friction constant
  $\zeta=0.5\mathrm{m}/\tau$.  Information between MD and CFD is
  exchanged after every fixed time interval $\Delta t_{c}$, with
  $\Delta t_{c}=n_{CFD} \Delta t_{CFD}= n_{MD}\delta t$. Here we
  typically used $\Delta t_c =\Delta t\simeq 0.03\tau$ and $\delta t =
  0.0003\tau=0.5$fs (small enough to recover O-H vibrational motion).
\label{water_3s}}
\end{figure}

\newpage

\begin{figure}[!ht]
\vspace{0.6cm}
\includegraphics[width=\linewidth]{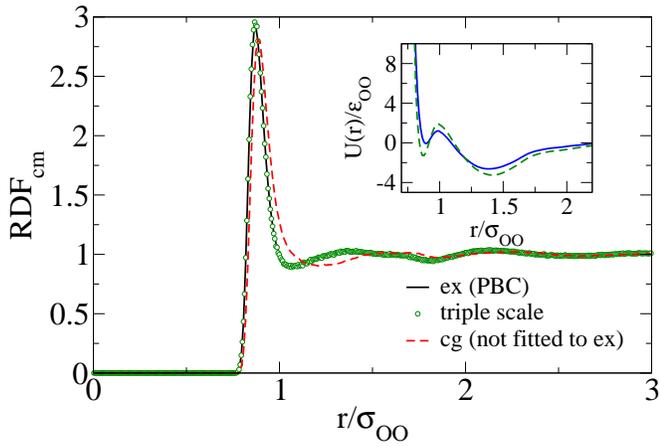}
\caption{Comparison between RDF$_{cm}$s obtained in a all-atom ($ex$
model) periodic boundary simulation of the flexible TIP3P water at ambient
conditions and within the MD region of a triple-scale simulation. At
the buffer we used a $cg$ model whose RDF$_{cm}$
(dashed line) was not fitted to the all-atom result.  The inset shows
the effective potentials $U^{cm}_{w=0}$ and $U^{cm}_{w=0.5}$ used in
the first protocol (see text), which correctly reproduce the all-atom
$RDF_{cm}$: two minima corresponding to the first and second hydration
shells of the liquid are observed.
\label{rdf}
}
\end{figure}

\newpage

\begin{figure}[!ht]
\includegraphics[width=\linewidth]{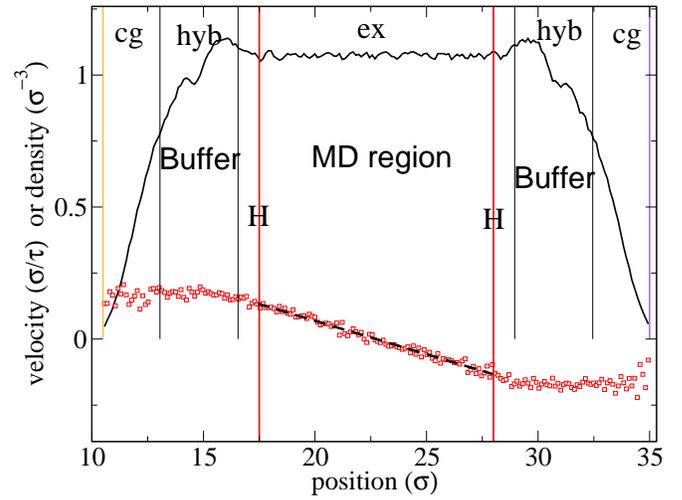}
\caption{Density profile (solid line) and velocity distribution (squares)
across the particle domain in a steady Couette flow (the dashed line is the expected
linear profile across the MD region).
\label{shear}
}
\end{figure}

\newpage

\begin{figure}[!ht]
\includegraphics[width=\linewidth]{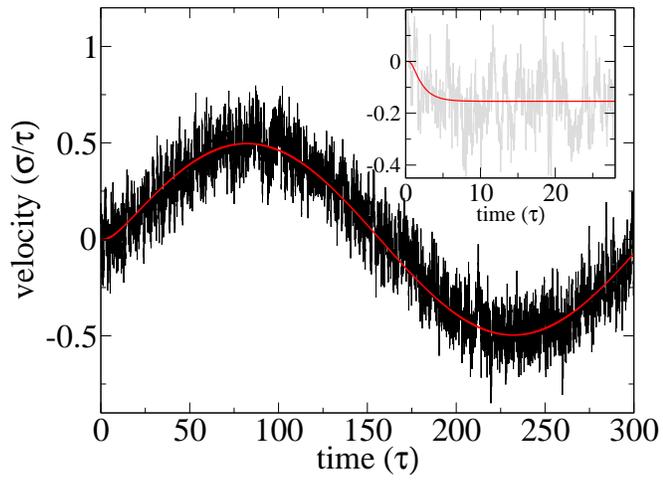}
\caption{Time evolution of the velocity at one finite volume cell within
the MD region in an oscillatory shear flow.
For comparison, the deterministic Navier-Stokes solution is shown in red lines.
The inset shows velocity in one MD cell at the start-up of a Couette flow.
\label{osc_shear}}
\end{figure}

\end{document}